\pdfoutput=1
\documentclass[sigconf]{acmart}

\usepackage{amssymb}
\usepackage{pifont}
\usepackage{url}

\usepackage{algorithm}
\usepackage{algpseudocode}

\usepackage{textcomp}
\usepackage{listings}
\usepackage{array,tabularx}
\usepackage{color}
\definecolor{myblue}{rgb}{0.2,0.4,0.8}
\definecolor{mygreen}{rgb}{0.3,0.7,0.3}
\definecolor{myred}{rgb}{0.8,0.4,0.2}
\definecolor{myyellow}{rgb}{0.7,0.7,0.3}
\definecolor{mygray}{rgb}{0.6,0.6,0.6}
\usepackage{ragged2e}
\newcolumntype{P}[1]{>{\RaggedRight\hspace{0pt}}p{#1}}
\usepackage{caption}
\usepackage{newfloat}
\DeclareFloatingEnvironment[
fileext=lol,
name=Listing,
placement={!ht}
]{listing}
\usepackage{subcaption}

\setcopyright{none} 
\acmConference[Anonymous Submission to ACM CCS 2018]{ACM Conference on Computer and Communications Security}{Due 8 May 2018}{Toronto, Canada}
\acmYear{2018}

\newcommand{\scbf}[1]{\vspace {0.05in}\noindent{\textbf{#1.}}}
\newcommand{\algorithmicbreak}{\textbf{break}}
\newcommand{\Break}{\State \algorithmicbreak}
\newcommand{\algorithmiccontinue}{\textbf{continue}}
\newcommand{\Continue}{\State \algorithmiccontinue}

\algrenewcomment[1]{\(\triangleright\) #1}
\algnewcommand{\LineComment}[1]{\State \(\triangleright\) #1}

\newcommand*\fixme[1]{\textbf{\color{red}#1}}

\begin{document}
\title{Sound Patch Generation for Vulnerabilities} 

\author{Zhen Huang}
\email{z.huang@mail.utoronto.ca}
\affiliation{University of Toronto}
\author{David Lie}
\email{lie@eecg.toronto.edu}
\affiliation{University of Toronto}

\begin{abstract}
Security vulnerabilities are among the most critical software defects in existence.  As such, they require patches that are correct and quickly deployed.  This motivates an automatic patch generation method that emphasizes both soundness and wide applicability.  To address this challenge, we propose Senx, which uses three novel patch generation techniques to create patches for out-of-bounds read/write vulnerabilities. Senx uses symbolic execution to extract expressions from the source code of a target application to synthesize patches.  To reduce the runtime overhead of patches, it uses \textit{loop cloning} and \textit{access range analysis} to analyze loops involved in these vulnerabilities and elevate patches outside of loops.  For vulnerabilities that span multiple functions, Senx uses \textit{expression translation} to translate expressions and place them in a function scope where all values are available to create the patch.  This enables Senx to patch vulnerabilities with complex loops and interprocedural dependencies that previous semantics-based patch generation systems cannot handle.

We have implemented a prototype using this approach. Our evaluation shows that the patches generated by Senx successfully fix 76\% of 42 real-world vulnerabilities from 11 applications including various tools or libraries for manipulating graphics/media files, a programming language interpreter, a relational database engine, a collection of programming tools for creating and managing binary programs, and a collection of basic file, shell, and text manipulation tools.  All patches that Senx produces are sound, and Senx correctly aborts patch generations in cases where its analysis will fall short.
\end{abstract}

\ccsdesc[500]{Security and privacy~Software security engineering}

\keywords{software vulnerabilitiy; automated patch generation; } 

\maketitle

\newcommand{\code}[1]{\texttt{#1}}

\begin{lstlisting} [float,floatplacement=T,label=lst:example_vulnerability,language=C,caption=A buffer overflow CVE-2012-0947 with a patch (prefixed with '+'). \vspace{-20pt},escapechar=$]
char *foo_malloc(x,y) {
	return (char *)malloc(x * y + 1);$\label{lst:malloc_call}$
}

// print a flattened array 
int foo(char *input) {
	// input format: RRCCDDD....DDD
	//	RR: number of rows
	//	CC: number of columns
	//	DDD...DDD: flattened array data
	//
	// benign input: 0203123456
	// output: 
	//		1 2 3
	//		4 5 6
	int size = strlen(input);
	char *p = input;
	int rows = extract_int(p);	
	p +=2;
	size -= 2;
	int cols = extract_int(p);
	p +=2;
	size -=2;
+	if ((double)(cols+1)*(size/cols) > $\label{lst:patch_start}$
+			rows * (cols + 1) + 1)
+		return -1; $\label{lst:patch_end}$
	char *output = foo_malloc(rows, cols + 1); $\label{lst:foo_malloc_call}$
	if (!output)
		return -1;
	bar(p, size, cols, output);
	printf("%s\n", output);
	free(output);
	return 0;
}

void bar(char *src,int size,int cols,char *dest) {
	char *p = dest;char *q = src;
	while (q < src + size)  {
		for (unsigned j = 0; j < cols; j++)  $\label{lst:inner_start}$
			*(p++) = *(q++);  $\label{lst:inner_end}$
		*(p++) = '\n';
	}
	*p = '\0';
}

\end{lstlisting}


\section{Introduction}
Fixing security vulnerabilities in a timely manner is critical to protect users from security compromises and to prevent vendors from losing user confidence. A recent study has shown that creating software patches is often the bottleneck of the process of fixing security vulnerabilities~\cite{Talos}.  As a result, an entire line of research inquiry into automated patch generation has arisen to try to address this challenge~\cite{PAR,SPR,LeakFix,GenProg,RSRepair,AE,APR-Plausibility,SemFix,Angelix,Prophet}. 

Automatic patch generation approaches broadly break down into two categories: search-based and semantics-based.  Search-based approaches try arbitrary code changes and use a battery of test cases to check whether any of the changes succeeded in fixing the bug~\cite{PAR,SPR, GenProg}.  Because various search-based techniques can generate arbitrary code changes, they are applicable to a wide variety of bugs, but the correctness of the generated patches can be uneven, and depend heavily on the comprehensiveness of the test cases~\cite{Monperruscriticalreviewautomatic2014,APR-Plausibility}.  In contrast, semantics-based techniques use code analysis (e.g. static analysis and symbolic execution) to produce patches that attempt to address the underlying bug rather than change the code just enough to satisfy the test cases~\cite{SemFix,Angelix}.  As a result, semantics-based techniques are more likely to produce correct patches, but are less applicable, since they are limited to cases where the code analysis is able to analyze the defective code.

When used to fix security vulnerabilities, the requirements that patches work correctly is even more pressing, as falsely believing that a vulnerability has been addressed when in fact it has not, may lead a user to disable other mitigating protections, such as removing configuration workarounds or firewall rules.  In this paper, we propose Senx, which guarantees that either the patch it produces addresses the vulnerability under all conditions (i.e. it is sound), or it does not produce a patch at all.  Despite this rigorous requirement, we find that using sound semantic analysis Senx is still able to produce patches for over 76\% of the vulnerabilities in a bug corpus composed of buffer overflows, integer overflows and memory corruption due to bad type-casts.  These vulnerabilities involve complex code structures that code analysis techniques traditionally find challenging.  For example, buffer overflows often involve complex loop structures.  In addition, formulating a check to test a memory access is within the memory-range of the data-structure, may require the patch generation system to synthesize interprocedural code if the allocation of the memory and the faulty memory access occur in different functions.

To concretely illustrate these challenges, consider the buffer overflow vulnerability CVE-2012-0947~\cite{CVE-2012-0947} from libav in Listing~\ref{lst:example_vulnerability}.  The vulnerability arises because the code copies user-provided data into a buffer whose allocation size is computed based on the number of rows and columns specified in the input, but the amount of data copied is based on the size of the data provided.  For reference, the correct patch is provided on lines~\ref{lst:patch_start}-~\ref{lst:patch_end}, which consists of checking if the amount of data to be copied (\texttt{(cols+1)$\times$(size/cols)}) by the nested loop in \texttt{bar} is greater than the size of the buffer (\texttt{rows$\times$(cols+1)+1}) allocated by \texttt{foo\_malloc}, in which case the patch returns an error to \texttt{foo}'s caller.  To generate this patch, Senx must correctly identify the pointer \texttt{p} used to write to the buffer \texttt{dest} in \texttt{bar} and infer the memory access range of \texttt{p} from the nested loops.  Further, Senx must detect that the allocation of \texttt{dest} is actually performed in another function \texttt{foo\_malloc} and symbolically compute its allocation size.  Finally, Senx must identify \texttt{foo} as the common caller of both \texttt{bar} and \texttt{foo\_malloc} and translate the expression for both the memory access range of \texttt{p} from \texttt{bar} and the allocation size of \texttt{dest} from \texttt{foo\_malloc} into the scope of \texttt{foo} so that the patch can be generated.

Senx accomplishes this with the introduction of three novel patch generation techniques. First, rather than try to statically analyze arbitrarily complex loops, Senx attempts to clone the loop code to be used in the patch predicate, but slicing out any code that may have side effects from the cloned loop so that the loop only computes the memory access range of the pointers in the loop.  We call this technique \textit{loop cloning}.  For loops where code that has side effects that cannot be safely removed -- for example, the loop execution depends on the result of a function call with side effects---Senx falls back on a symbolic analysis technique we developed, called \textit{access range analysis}.  Finally, to identify and place the patch in a function scope where all expressions required in the predicate are available, Senx uses \textit{expression translation}, which uses the equivalence between function arguments and parameters to generate a set of equivalent expressions.  This enables Senx to generate patches where the defective code is spread across multiple functions.  This overcomes a limitation of all previous semantics-based patch generate systems that we are aware of, which can only generate patches for defects that are enclosed entirely within in a single function~\cite{PAR, SemFix, Angelix, SPR}.

In our experiments, Senx generates 32 correct patches out of 42 vulnerabilities, and in the remaining 10 cases, correctly detects that it will not be able to generate a patch and aborts instead of generating a patch that does not completely fix the vulnerability.  

This paper makes the following main contributions:
\begin{itemize}
	\item We describe the design of Senx, an automatic patch generation system for buffer overfow, integer overflow and bad cast vulnerabilities using three novel program analysis techniques: loop cloning, access range analysis and expression translation.
	\item Senx uses a symbolic ISA that is optimized for converting an IR used for symbolic execution back to source code.  This enables Senx to generate source code patches.  Particularly, Senx's ISA supports the synthesis of source code containing array indices and C/C++ structs and classes.  
	\item We prototype Senx on top of the KLEE symbolic execution engine and evaluate it on a corpus of 42 vulnerabilities across 11 popular applications, including PHP interpreter, sqlite database engine, binutils utilities for creating and managing binary programs, and various tools or libraries for manipulating graphics/media files.  Senx generates correct patches in 32 of the cases and aborts the remainder because it is unable to determine semantic correctness in the other cases.  The evaluation demonstrates that all three techniques are required to generate patches, and that failure to find a common function scope in which to place a patch is the most frequent reason for failure.
\end{itemize}

The structure of this paper is as follows. In Section~\ref{sec:definitions}, we define and characterize the problem Senx addresses. Section~\ref{sec:senx_design} and Section~\ref{sec:implementation} describe the design and implementation of Senx respectively. We present evaluation results in Section~\ref{sec:senx_evaluation} and discuss related work in Section~\ref{sec:related_work}. Finally we conclude in Section~\ref{sec:senx_conclusion}.

\section{Problem Definition}\label{sec:definitions}

We begin by defining what a Senx patch is, what guarantees that a Senx patch provides, and then the vulnerability domain, i.e. vulnerability types that Senx can currently address.

\subsection{Patch} 
For a given vulnerability, a Senx patch fixes the corresponding \textit{vulnerability condition}~\cite{VulnerabilitySignature}, which evaluates to true if and only if the vulnerability will be triggered. A Senx patch can take one of two forms: a) detects if a vulnerability condition is true and if so, raises an error to direct program execution away from the path where the vulnerability resides (we call this a check-and-error patch); b) prevents a vulnerability condition from ever becoming true (we call this a repair patch).  

\scbf{Soundness} Because the patches that Senx generates directly address the vulnerability condition, it prevents the vulnerability from being triggered under all conditions. Senx only generates patches if the vulnerability falls into one of the vulnerability classes in its \textit{vulnerability domain} and the vulnerable code meets a number of requirements that Senx requires to guarantee the soundness of the patch.  If all these conditions are met, Senx will generate a sound patch, and if they are not, Senx does not generate a patch at all.

\subsection{Vulnerability domain}\label{sec:vuln_domain}
Senx is designed to generate patches for specific, common vulnerability types in its vulnerability domain.  Each such vulnerability class is defined by a vulnerability condition.  We describe the three vulnerability classes currently in Senx's vulnerability domain:

\scbf{Buffer overflow} A buffer overflow occurs when series of memory accesses traversing a buffer in a loop crosses from a memory location inside the buffer to a memory location outside of the buffer.  We use the term \textit{buffer} in the broad sense to refer to either a bounded memory region (such as a struct or class object) or an array. The memory access can be the result of an array dereference or pointer dereference.  Senx handles both the case when the memory access exceeds the upper range of the buffer and the case when the memory access falls below the lower range (sometimes called a buffer underflow). The vulnerability condition for buffer overflow evaluates whether the range of memory addresses that code within a loop accesses exceeds either the upper or lower range of the buffer.

\scbf{Bad cast}  A bad cast occurs when a memory access is computed off a base pointer, and exceeds the upper bound of the object the base pointer points to. It is caused by a cast of a pointer to an incorrect object type. The memory access can be an access to a struct field or a class attribute and the struct or class can be nested. The vulnerability condition for bad cast evaluates whether the memory address targeted by a memory access exceeds the upper bound of the object pointed to by the base pointer used to compute the memory address. 


\scbf{Integer overflow} An integer overflow occurs when an integer is assigned a value larger or smaller than can be represented. This manifests as a large value that is increased and becomes a small value, or a small value that is decreased and becomes a large value. Senx addresses integer overflows whose result is zero and used as the allocation size of a buffer, because any subsequent memory access to a buffer of zero byte will fall outside the buffer. The vulnerability condition for such integer overflow evaluates whether the size specified for a buffer allocation is zero and is the result of an integer overflow.

As a prototype, we have started with only these three vulnerability classes.  Nevertheless, they are representative enough to capture a good percentage of CVE vulnerabilities---we conducted an informal analysis and found that roughly 10\% of CVE vulnerabilities fall into one of these classes.  We believe the principles behind Senx can be extended to other vulnerability classes, which we intend to do in future work.

\section{Design}\label{sec:senx_design}



\subsection{Overview}
To generate a patch, Senx requires an input that can trigger the target vulnerability, which it runs on the target program on a symbolic execution engine.  This execution is the first of five steps to generate a patch: vulnerability detection, vulnerability identification, predicate generation, patch placement, and patch synthesis.

First, during \textit{vulnerability detection}, Senx symbolically executes the application with a vulnerability-triggering input and extracts expressions for all variables used or defined during the execution.  While many systems use symbolic execution to extract path constraints and generate inputs for program coverage, Senx only executes the single path that results from the vulnerability-triggering input and focuses on extracting symbolic expressions in such a way that a source code patch can be synthesized.  Senx determines a \textit{vulnerability point}~\cite{VulnerabilitySignature} at which the application meets one of the vulnerability conditions in its vulnerability domain.  Senx will abort patch generation if any variable used during the execution is the result of an operation that cannot be expressed by Senx's expression builder, which is described in Section~\ref{sec:symbolic_expression_builder}---if the expressions are incomplete, Senx cannot guarantee the soundness of the generated patch.  If no vulnerability point can be detected, Senx will also abort patch generation.

Second, during \textit{vulnerability identification}, Senx uses the results of the execution to classify the vulnerability into one of the three classes given in Section~\ref{sec:vuln_domain} using their respective vulnerability conditions.  During execution, Senx also performs a reachability analysis to ensure that the executed path covers all reaching definitions for variable expressions in the vulnerability condition.  If statically there exists a reaching definition that wasn't on the executed path, then the expression for one of the variables in the vulnerability condition may be incomplete and the patch produced may be unsound, and Senx will abort patch generation under these circumstances.  To ensure that Senx statically captures all reaching definitions, Senx's reachability analysis is enhanced with DSA pointer alias analysis~\cite{DSA} to ensure that there are no aliases that might alter a reaching definition.  The pointer analysis returns a confidence category that can be either ``must alias'', ``must not alias'' or ``may alias''.  If any pointer ``must alias'' or ``may alias'' with a variable in the vulnerability condition, Senx also aborts patch generation.


Third, during \textit{predicate generation}, using the vulnerability class and associated condition, Senx will generate a predicate that will be used to create a patch for the vulnerability.  We describe the predicate generation process in more detail in Section~\ref{sec:predicate_generation}.  For each vulnerability class, we will also describe the set of requirements that must be met to guarantee soundness.

Fourth, during \textit{patch placement}, Senx finds a location to place a patch where all the variables in the predicate are in program scope.  In some cases, the predicate may contain variables from different function scopes, so that a single statement generated using those variables could not be evaluated at any one place in the program.   For example, in Listing~\ref{lst:example_vulnerability}, the size of buffer \code{dest} is defined by an expression over the variables \code{x} and \code{y} in \code{foo\_malloc}, which are not available in the scope of function \code{bar} where the loop that overflows occurs.  In these cases, Senx uses \textit{expression translation} to translate those variables into expressions that are in the scope of a common caller or callee of the functions.  In Listing~\ref{lst:example_vulnerability}, Senx recognizes that function \code{foo} is a common caller of both \code{foo\_malloc} and \code{bar} and translates all the terms of the predicate into expressions over variables available in the scope of \code{foo}.  We describe expression translation in more detail in Section~\ref{sec:symbolic_expression_translation}.  Senx will abort patch generation if expression translation fails.

\begin{table*}[t]
	{\footnotesize	
		\caption{Rules for building expressions.}
		\label{tbl:symbolic_expression_rules}
		
		\vspace{-12pt}
		\centering
		\begin{tabular}{|l|l|l|l|}
			\hline
			\textbf{Instruction} & \textbf{Semantic} & \textbf{Rule to build expression} & \textbf{Description}\\
			\hline
			$val$ = Load $var$ & $val \leftarrow var$ & RHS := getExpr$(var)$ & read from $var$ \\
			$val$ = Load $*p$ & $val \leftarrow *p$ & RHS := makeDeref(getExpr$(p)$) & read via pointer $p$ \\
			Store $var, v1$ & $p \leftarrow v1$ & RHS := getExpr$(v1)$, LHS := $var$ & write $v1$ to $var$ \\
			Store $*p, v1$ & $*p \leftarrow v1$ & RHS := getExpr$(v1)$, LHS := makeDeref($p$) & write via pointer $p$ \\
			$val$ = GetElement $var, field$ & $val \leftarrow$ StructOp($var, field$) & RHS := makeStructOp($var, field$) & read from a struct field \\
			$val$ = GetElement $var, index$ & $val \leftarrow$ ArrayOp$(var, index)$ & RHS := makeArrayOp$(var, index)$ & read from an array element \\
			$val$ = BinOp $v1, v2$ & $val \leftarrow$ BinOp$(v1,v2)$ & RHS := makeBinOp(getExpr($v1$), getExpr($v2$)) & binary operations \\
			$val$ = CmpOp $v1, v2$ & $val \leftarrow$ CmpOp$(v1,v2)$ & RHS := makeCmpOp(getExpr($v1$), getExpr($v2$)) & comparisons \\
			$val$ = Allocate $size$ & $val \leftarrow$ Allocate$(size)$ & RHS := getName($val$) & allocate a local variable \\
			Branch $label$ & $PC \leftarrow label$ & N/A & unconditional branch \\
			Branch $cond, label1, label2$ & $PC \leftarrow label1$ if $cond$ & N/A & conditional branch \\
			& $PC \leftarrow label2$ if $\neg cond$ & N/A &\\
			$val$ = Call $f(a, \dots)$ & $val \leftarrow f(a, \dots)$ & RHS := makeCall(getName($f$), getExpr($a$), \dots) & call function $f$ with $a, \dots$\\
			Ret $v1$ & $val \leftarrow v1$ \&\& & RHS := getExpr($v1$) \&\& & return \textit{v1} to caller\\
			& $caller.val \leftarrow v1$ & $caller$.RHS := getExpr($v1$) & \\

			\hline
		\end{tabular}
		\vspace{-10pt}
	}
\end{table*}

Finally, during \textit{patch synthesis},  Senx generates the patch code.  The expressions that Senx extracts during vulnerability detection are specially designed for easy translation back into C/C++ source code.  For example, Senx maintains special information that allows it to synthesize field dereferences of structs (even nested structs) or array dereferences.  Currently the only vulnerability class for which a repair patch can be generated is an integer overflow where the overflow occurred as a result of an arithmetic operation whose result type does not match the type of the variable the result was being stored into.  In this case, a patch that casts the result into the type of the variable avoids the overflow and is inserted into the code.  In all other cases, Senx generates a check-and-error patch that checks the patch predicate and calls the error handling code if the predicate evaluates to true. Senx uses Talos~\cite{Talos} to find and select the error handling code to call.  Senx will abort patch generation for check-and-error patches if Talos cannot determine suitable error-handling code to call.



\subsection{Expression Builder}\label{sec:symbolic_expression_builder}

We base our expression builder on KLEE~\cite{KLEE}, but as we mentioned earlier, Senx uses its own symbolic representation as KLEE's representation does not store enough information to easily translate expressions back into source code to construct patches.  As a result, we design our own symbolic representation using a set of pseudo-instructions listed in Table~\ref{tbl:symbolic_expression_rules}.

The instructions include \texttt{Load} and \texttt{Store} memory access instructions, \texttt{BinOp} binary operations such as arithmetic operations and \texttt{CmpOp} comparison operations such as $>$ and $\geq$, \texttt{StructOp} struct operations that access a field of a struct, \texttt{ArrayOp} array operations that access an element of an array, \texttt{Allocate} for local variable allocation, \texttt{Branch} for unconditional and conditional branches, \texttt{Call} for function calls and \texttt{Ret} for function returns. Each instruction can have an optional label denoted as \texttt{label}.  The decompilation uses a program counter that points to the current instruction, which is referred to as \texttt{PC} in the table. For each instruction, Senx represents it using the corresponding semantic in the ``Semantic'' column.  The results of these operations are stored in Single Static Assignment (SSA) form such that each instruction instance has a unique variable associated with it.   The execution does not distinguish between registers and memory.

Later, during patch synthesis, source code is generated using the rules defined in the ``Rule to build expression'' column.  Each semantic expression is of the form LHS := RHS, where LHS and RHS are the left side and the right side of an assignment respectively. An explicit LHS is used only for Store instructions.  The LHS for all other instructions is the SSA value associated with each instruction.  The expression generation maintains a call stack so that each \textit{Ret} instruction sets a value in its caller, denoted by \texttt{caller}, with the return value.  A good deal of the heavy lifting for expression generation is done by helper functions as described in Table~\ref{ap:tbl:SEB_operations}.  Each of these functions generates an expression according to their description. For example, \texttt{makeDeref("p")} returns "\texttt{*p}", where $*$ represents pointer dereference.  In keeping with SSA, the expressions generated for an instruction are stored along with the instruction.  In this way, the expressions associated with an instruction can easily be retrieved by referring to the instruction.


\begin{table}
	\caption{Operations performed by expression builder.}
	\label{ap:tbl:SEB_operations}	
	\vspace{-8pt}
	\begin{center}
		\begin{tabular}{|p{0.75in}|p{2.2in}|}
			\hline
			\textbf{Operation} & \textbf{Description} \\
			\hline
			getExpr & get the expression associated with an \\
			& instruction or the name of a variable \\
			getName & get the name of a variable \\
			makeDeref & build an expression to denote dereference \\
			makeBinOp & build an expression to denote a binary \\
			& operation including \textit{arithmetic} operations, bitwise \textit{logic} operations, and bitwise \textit{shift} operations \\
			makeCmpOp & build an expression to denote a comparison \\
			& including $<,>,=,\ne, \geq, \leq$ \\
			makeStructOp & build an expression to denote an access \\
			& to a struct field directly or via a pointer \\
			makeArrayOp & build an expression to denote an access \\
			& to an array element  \\
			makeCall& build an expression to denote a function call \\
			& including the name of the function and all the arguments\\
			\hline 
		\end{tabular}
		\vspace{-20pt}
	\end{center}
\end{table}

%

\scbf{Complex Data Types} Because the patch generated by Senx is in the form of the source code of a target program, the expressions must conform to the proper language syntax of the program.

Expressions for simple data types such as char, integer, or float, are generated in a rather straightforward way.  However, expressions for complex data types such as C/C++ structs and arrays are more challenging.  For example, a field of a struct must be attached to its parent object, and the generated syntax changes depending on whether the parent object is referenced using a pointer or with a variable holding the actual object.  Arrays and structs can also be nested and the proper syntax must be used to denote the level of nesting relative to the top level object.

To address the challenge, Senx's symbolic ISA has the \texttt{GetElement} instruction, which reads a field from a struct or an element from an array. The expression builder leverages the \texttt{GetElement} instructions and debug symbols that describe the ordered list of struct fields to construct expressions denoting access to complex data types including arrays and structs. \texttt{GetElement} is overloaded, but since the expression builder maintains the data type of each variable, it calls the appropriate version based on the type passed in \texttt{var}.  To generate valid C/C++ code for a symbolic expression, it retrieves the variable expression associated with \texttt{var}.  If \texttt{var} is an array, it uses the helper function \texttt{makeArrayOp}, which recursively generates code associated with the \texttt{index} argument.  If \texttt{var} is a struct, it calls the helper function \texttt{makeStructOp} , which returns the name of the field in the struct.  To determine whether an access to the struct is via a pointer or directly to an object, it checks whether \texttt{var} is a result of a \texttt{Load} instruction or not, and generates the expression accordingly.

In order to build expressions for complex data access involving nested complex data types, both \texttt{makeArrayOp} and \texttt{makeStrutOp} use the expression for the variable \texttt{var}, which can be the result of a previous \texttt{Load} instruction or \texttt{GetElement} instruction. In this way, expressions for complex data access such as \texttt{foo$\rightarrow$f.bar[10]}, where \texttt{foo} is a pointer to a struct that has a field \texttt{f} and \texttt{bar} is an array belonging to \texttt{f}, can be constructed.

\subsection{Predicate Generation}\label{sec:predicate_generation}

Predicate generation is specific to the vulnerability condition that Senx identifies.  We discuss in detail how Senx generates predicates for each vulnerability class in its vulnerability domain.  

\scbf{Integer Overflow} Since Senx has already identified the arithmetic operation variables that will overflow in the previous step, it is straightforward for Senx to generate a predicate that will check the values of the variables before the arithmetic operation to see if they will generate an overflow.  In this case, given that the soundness requirements have already been satisfied in the vulnerability identification stage, the expressions extracted during that stage will be sound, so the predicate that performs a check over those expressions will also be sound.  Once the predicate is generated, Senx determines whether it should generate a repair patch (with a cast) or a check-and-error patch.  So long as the extracted expressions are sound, the generated predicate is sound.

\scbf{Buffer Overflow and Bad Cast} Senx fixes both of these vulnerability classes with check-and-error patches.  The predicates produced for both vulnerability classes are similar in structure and consist of a comparison of a memory access with an allocation range.  For bad casts, the memory access is a single \textit{memory access point}, while for buffer overflows, it is a series of accesses performed in the loop that forms a \textit{memory access range}.  In both cases, it is compared with the size of the object being accessed, which is extracted when the object is allocated. We refer to the size as the object's \textit{allocation range}.  With this, Senx generates a predicate of the form: $(mem\_access\_upper > alloc\_upper  || mem\_access\_lower < alloc\_lower)$.  In the case where it is a memory access point instead of a memory access range, both $mem\_access\_upper$ and $mem\_access\_lower$ are equivalent.

To compute the memory access point for a bad cast, Senx uses the expressions extracted to compute the offset of the struct field or class attribute being accessed and generates an expression that adds the offset to the base pointer being used in the access. We define the base pointer as the pointer that points to the start address of the struct or class object to which the field or attribute being accessed belongs. If the field or attribute is a direct member of the struct or class, i.e. not a member of a nested struct or class, the base pointer is simply the pointer extracted from the expressions. Otherwise, we perform interprocedural dataflow analysis to find the base pointer from which the extracted pointer is derived.  As in the case of the integer overflow, so long as soundness requirements for the expression builder are met, the extracted memory access point is also sound.

The allocation range is computed in a similar way.  Senx identifies the point where the buffer is allocated using expressions extracted by the expression builder and examines the argument passed to the allocation function (i.e. \texttt{malloc} or some variant). With this, Senx can calculate the upper and lower bounds of the object being accessed.  This computation is sound if the expressions extracted are sound and the memory allocation is correctly identified, which is true if the application only uses standard memory allocation functions that Senx knows about.  Our Senx prototype currently only supports standard libc and C++ memory allocation functions.

Finally, computing a memory access range over a loop is more complex because loops may execute for a variable number of iterations.  To compute the memory access range, Senx uses two complementary loop analysis techniques, loop cloning and access range analysis. Both loop cloning and access range analysis take as input a function \texttt{F} in the target program and an instruction \texttt{inst} that performs the faulty access in the buffer overflow, and returns the symbolic memory access range \texttt{$[A_1, A_n]$} of \texttt{inst}.  This symbolic access range can then be compared with the allocation range in the predicate.

\label{sec:loop_cloning}
\scbf{Loop Cloning}  The key idea of loop cloning is to produce new code that can be called safely at runtime to return the access range without causing any side-effects, i.e. changing program state or affecting program input/output. The new code is constructed from existing code, referred to as \textit{cloning}, and will be called at a location where the allocation range is available so that the access range returned by the new code can be compared against the allocation range.


%
%

Because the patch must be inserted into a function where both the access range and allocation range are available, loop cloning first searches on the call chain that leads to \texttt{F} to find such a function. The search starts from the immediate caller of \texttt{F} and stops at the first function \texttt{F$_p$} in which the allocation range is available. 

If no such function can be found, Senx will not be able to generate a patch. If such a function is found, loop cloning then clones each function \texttt{F$_i$} along the call chain from \texttt{F} until \texttt{F$_p$} into the new code that returns the access range. As a result, each \texttt{F$_i$} is either a direct or indirect caller of \texttt{F} or is \texttt{F} itself. 

Loop cloning needs to satisfy two requirements: 1) \texttt{F} must compute the access range and pass the access range to its caller; 2) any direct or indirect caller of \texttt{F} must pass the access range that it receives from its callee upwards to the next function along the call chain. Each \texttt{F$_i$} is cloned using the following steps.

\begin{enumerate}
	\item Loop cloning clones the entire code of \texttt{F$_i$} into \texttt{F$_i$\_clone}.
	\item Using program slicing, it removes all statements that are not needed in order to computer the access range or pass the access range to \texttt{F$_p$}.  If \texttt{F$_i$} is \texttt{F}, it retains statements on which the execution of \texttt{inst} is dependent. If \texttt{F$_i$} is a direct or indirect caller of \texttt{F}, it retains statements on which the call to \texttt{F} is dependent.
	\item It changes the return type of \texttt{F$_i$\_clone} to \texttt{void} and removes any return statement in \texttt{F$_i$\_clone}. 
	\item It adds two output parameters \texttt{start} and \texttt{end} to \texttt{F$_i$\_clone}. If \texttt{F$_i$} is \texttt{F}, it inserts statements immediately before the (nested) loops to copy the initial value of the pointer or array index used in the faulty access into \texttt{start}, and statements immediately after the loops to copy the end value of such pointer or array index into \texttt{end}. If \texttt{F$_i$} is a caller of \texttt{F}, it changes the call statement to include the two output parameters in the list of call arguments.
\end{enumerate}

After cloning each \texttt{F$_i$}, loop cloning inserts a call to the last cloned function into \texttt{F$_p$}, which returns the access range in \texttt{start} and \texttt{end}. Subsequently a patch will be synthesized to leverage the returned access range.

\begin{lstlisting}  [float, floatplacement=T,label=lst:example_complex_loop,language=C,caption=A complex loop involving a complex loop exit condition and multiple updates to loop induction variable on multiple execution paths. \vspace{-18pt}]
int decode(const char *in, char *out) {
	int i;
	char c;
	i = 0;
	while ((c = *(in++)) != '\0') {
		if (c == '\1')
			c = *(in++) - 1;	
		out[i ++] = c;
	}
	return i;
}

char* udf_decode(const char *data, int datalen) {
	char *ret = malloc(datalen);
	if (ret && !decode(data+1, ret)) {
		free(ret);
		ret = NULL;
	}
	return ret;
}
\end{lstlisting}

\begin{lstlisting} [float, floatplacement=T,label=lst:example_cloned_loop,language=C,caption=A cloned and sliced loop that no longer contains any statements that have side-effects and returns the number of iterations. Statements prefixed with '+' are added or modified by Senx to count and return the number of loop iterations. \vspace{-18pt}]
+	void decode_clone(const char *in, char *out, char **start, char **end) {
		char c;
+		*start = in;
		while ((c = *(in++)) != '\0') {
			if (c == '\1')
				c = *(in++) - 1;	
		}
+		*end = in;
	}

	char* udf_decode(const char *data, int datalen) {
		char *ret = malloc(datalen);
+		char *start, *end;
+		decode_clone(data+1, ret, &start, &end);
		if (ret && !decode(data+1, ret)) {
			free(ret);
			ret = NULL;
		}
		return ret;
	}
\end{lstlisting}

To see how loop cloning works, consider the example in Listing~\ref{lst:example_complex_loop}, which presents a loop adapted from a real buffer overflow vulnerability CVE-2007-1887 \cite{CVE-2007-1887} in PHP, a scripting language interpreter. The buffer overflow occurs in function \texttt{decode}. The loop features a complex loop exit condition and multiple updates to loop induction variable \texttt{in} that depend on the content of the buffer that \texttt{in} points to. The result of loop cloning is shown in Listing~\ref{lst:example_cloned_loop}.  Loop cloning is invoked with \texttt{decode} as \texttt{F}, and the faulty access at line 5 as \texttt{inst}. It first finds that function \texttt{udf\_decode} is on the call chain to \texttt{decode} and in which the allocation range is available. Because \texttt{udf\_decode} directly calls \texttt{decode}, it only needs to clone \texttt{decode}. It then clones function \texttt{decode} into \texttt{decode\_clone}, after which it applies program slicing to \texttt{decode\_clone} with line 5 and variable \texttt{c} and \texttt{in} that are accessed at line 5 as the slicing criteria.  \texttt{decode} also has a potential write buffer overflow at line 8, but in this example, we focus on generating a predicate that will check whether \texttt{in} can exceed the end of the buffer it points to. The program slicing uses a backward analysis and removes all statements that are irrelevant to the value of \texttt{c} and \texttt{in} at line 5, including line 2, 4 and 8.  After program slicing, it changes the return type of \texttt{decode\_clone} into \texttt{void} and removes all return statements. And it adds two output parameters \texttt{start} and \texttt{end} to the list of parameters of \texttt{decode\_clone}.  Then it inserts a statement at line 3 to copy the initial value of \texttt{in} to \texttt{start} before the loop and a statement at line 8 to copy the end value of \texttt{in} to \texttt{end} after the loop. Finally it inserts into function \texttt{udf\_decode} a call to \texttt{decode\_clone} at line 14 and a statement to declare \texttt{start} and \texttt{end} at line 13. 


To satisfy the soundness requirements, the new code produced by loop cloning must not have any side-effect of changing program state or affecting program input/output. If the produced code has side-effect, Senx will abort loop cloning.

\scbf{Access Range Analysis}\label{sec:access_range_analysis} Using LLVM's built-in loop canonicalization functionality~\cite{LLVM_LOOP_CANONICALIZATION}, access range analysis computes the access range of normalized loops. Loop canonicalization seeks to convert the loop into a standard form with a pre-header that initializes the loop iterator variable, a header that checks whether to end the loop, and a single backedge.  Extracting the access range for a single loop in this way is straightforward.  The main difficulty is extending this to handle nested loops.


Access range analysis is implemented for nested loops using the algorithm described in Algorithm~\ref{alg:access_range_analysis}. It analyzes the loops enclosing a memory access instruction \texttt{inst} in function \texttt{F}, starting with the innermost loop and iterating to the outermost, accumulating increments and decrements on the loop induction variables including the pointer used by \texttt{inst}. 

Since the loop in \texttt{bar} of Listing~\ref{lst:example_vulnerability} can be normalized, we use it as an example of how Algorithm~\ref{alg:access_range_analysis} can be applied to a nested loop.  So \texttt{F} is \texttt{bar} and \texttt{inst} is the memory write using pointer \texttt{p} at line 42. For each loop, it first retrieves the loop iterator variable and the bounds of it by calling helper function \texttt{find\_loop\_bounds}, and the list of induction variables of the loop along with the $update$ to each of them, which we refer to as the fixed amount that is increased or decreased to an induction variable on each iteration of the loop, by calling another helper function \texttt{find\_loop\_updates}.  In our example, we have $iter=\texttt{j}, initial=0, end=\texttt{cols}$ and $\texttt{j} \mapsto 1, \texttt{p} \mapsto 1, \texttt{q} \mapsto 1$ in $updates$ for the innermost \texttt{for} loop from lines~\ref{lst:inner_start}-\ref{lst:inner_end}.

Algorithm~\ref{alg:access_range_analysis} then symbolically accumulates the update to each induction variable to a data structure referred to by $acc$, which maps each induction variable to an expression denoting the accumulated update to the induction variable. As for the example, it will store $\texttt{j} \mapsto 1, \texttt{p} \mapsto 1, \texttt{q} \mapsto 1$ into $acc$ for the innermost for loop.  After that, it synthesizes the expression to denote the total number of iterations for the loop. At line 16 of the algorithm, we will have $count=\texttt{cols}$ which is simplified from \texttt{(cols-0)/1}.

\begin{algorithm} [tb]
	\caption{Finding the access range of a memory access.}\label{alg:access_range_analysis}
	\begin{algorithmic}
	\Require $F$: a function \\ \hspace*{\algorithmicindent} $inst$: a memory access instruction in $F$
	\Ensure $acc\_initial$: initial address acccessed by $inst$ \\ \hspace*{\algorithmicindent} $acc\_end$: end address accessed by $inst$
	\begin{algorithmic}[1]
		\Procedure{analyze\_access\_range}{}
		\LineComment{$acc$: accumulated updates to induction variables}
		\State $acc \gets \emptyset$
		\State $innermost\_loop \gets innermost\_loop(inst)$
		\State $outermost\_loop \gets outermost\_loop(inst)$
		\State $visited \gets \emptyset$
		\For{$l \in {[innermost\_loop, outermost\_loop]}$}
		\State ${iter, initial, end}$ $\gets$find\_loop\_bounds$(F, l)$
		\State ${updates, visited} \gets$find\_loop\_updates$(l, visited)$
		\LineComment{Symbolically add up induction updates}
		\For{$var,upd \in updates$}
		\State $acc\{var\} \gets$ sym\_add$(acc\{var\},upd)$
		\EndFor
		\LineComment{Symbolically denote the number of iterations of $l$ as $count$}
		\State $upd\_iter \gets updates\{iter\}$
		\State $count\gets$sym\_div(sym\_sub$(end,initial),upd\_iter))$
		\LineComment{Symbolically multiply induction updates by the number of iterations of $l$}
		\For{$var,upd \in acc$}
		\If{$\neg$is\_initialized\_in\_last\_loop($var$)}
		\State $acc\{var\} \gets$sym\_mul$(acc\{var\}, count)$
		\EndIf
		\EndFor
		\EndFor
		\State $ptr \gets$get\_pointer$(inst)$
		\State $first\_inst \gets$loop\_head\_instruction$(outermost\_loop)$
		\LineComment{Find the definition of $ptr$ that reaches $first\_inst$}
		\State $acc\_initial \gets$reaching\_definition$(F, first\_inst, ptr)$
		\State $acc\_end \gets$sym\_add$(acc\_initial, acc\{p\})$
		\State \textbf{return} ${acc\_initial, acc\_end}$
		\EndProcedure
	\end{algorithmic}
	\end{algorithmic}
\end{algorithm}

Having the total number of iterations, it multiplies the accumulated update for each induction variable by the total number of iterations. So $acc$ will have $\texttt{j} \mapsto \texttt{cols}, \texttt{p} \mapsto \texttt{cols}, \texttt{q} \mapsto \texttt{cols}$ after the loop from line 18 to 22 in Algorithm~\ref{alg:access_range_analysis}.

Once this is done, it moves on to analyze the next loop enclosing $inst$, which in Listing~\ref{lst:example_vulnerability} is the while loop enclosing the inner for loop. As a consequence, we will have $iter=\texttt{q}, initial=\texttt{src}, end=\texttt{src+size}$ and $\texttt{p} \mapsto 1$ in $updates$ at line 10 of the algorithm, $\texttt{j} \mapsto cols, \texttt{p} \mapsto cols+1, \texttt{q} \mapsto \texttt{cols}$ in $acc$ and $count=\texttt{size/cols}$ at line 17 of the algorithm, and finally $\texttt{j} \mapsto \texttt{cols}, \texttt{p} \mapsto \texttt{(cols+1)*(size/cols)}, \texttt{q} \mapsto \texttt{size}$ in $acc$. Note that the algorithm will not multiply the number of iterations of the loop to \texttt{j} because \texttt{j} is always initialized in the last analyzed loop, the innermost for loop.

After analyzing all the loops enclosing $inst$, the algorithm gets the pointer $ptr$ used by $inst$ and performs reaching definition dataflow analysis to find the definition that reaches the beginning of the outermost loop. As for the example, we will have $ptr=\texttt{p}$ and the assignment \texttt{p=dest} at line 39 of \texttt{bar} as the reaching definition for \texttt{p}. From this reaching definition, it extracts the initial value of \texttt{p}, $acc\_initial=\texttt{dest}$. Finally it gets the end value of \texttt{p}, $acc\_end=\texttt{dest+(cols+1)*(size/cols)}$ by adding the initial value \texttt{dest} to the accumulated update of \texttt{p}, \texttt{(cols+1)*(size/cols)} from $acc$. Hence it returns $[\texttt{dest,dest+(cols+1)*(size/cols)}]$ as the expressions denoting the access range $[A_1,A_n]$. 

Our access range analysis can be considered as a form of pattern-based loop analysis~\cite{DetectLoopBounds} with several differences. On one hand, access range analysis aims to derive the number of loop iterations as an expression involving program variables and/or constants, while pattern-based loop analysis aims to derive the number of loop iterations as a constant. On the other hand, access range analysis requires loops be normalized to fit the pattern required by pattern-based loop analysis, and relies on loop canonicalization to normalize loops. 

To satisfy the soundness requirements, access range analysis only works on loops that can be successfully normalized. If the loops cannot be normalized, Senx will use loop cloning instead of access range analysis .

\subsection{Expression Translation}\label{sec:symbolic_expression_translation}

Because the patches Senx generates are source code patches, the predicate of the patch must be evaluated in a single function scope. However, sometimes the allocation range is computed in one function scope, while the memory access range is computed in a different function scope.  So the expression denoting the allocation range and the expression denoting the memory access range are not both valid in a single function scope. To make the expressions both valid in a single function scope, one possible solution is to send the expression valid in a source function scope as a call argument to a destination function scope where the expression is not valid. This approach requires adding a new function parameter to the destination function, and adding a corresponding call argument at every call site of the destination function. We decided not to use this approach because it requires code changes to any function on the call chain from the source function to the destination function. In addition, unrelated functions that call any of the changed functions will also have to be changed, resulting in a very intrusive patch. 

\textit{Expression translation} solves this problem by translating an expression $exp_s$ from the scope of a source function $f_s$ to an equivalent expression $exp_d$ in the scope of a destination function $f_d$. It does not require adding new function parameters or call arguments like the aforementioned solution. Senx uses expression translation to translate both the buffer size expression and memory access range expression into a single function scope where the predicate will be evaluated.  We call this process \textit{converging} the predicate.

At a high level, expression translation can be considered as a form of lightweight function summarization~\cite{FunctionSummarization}. While function summarization establishes the relations between the inputs to a function and the outputs of a function, expression translation establishes the relations between the inputs to a function and a subset of the local variables of the function. It works by exploiting the equivalence between the arguments that are passed into the function by the caller and the parameters that take on the argument values in the scope of the callee.  Using this equivalence, expression translation can iteratively translate expressions that are passed to function invocations across edges in the call graph.  Formally, expression translation can converge the comparison between an expression $exp_a$, the memory access range expression in $f_a$, and $exp_s$, the buffer size expression in $f_s$ iff along the set of edges $\mathbb{E}$ connecting $f_a$ and $f_s$ in the program call graph, an expression equivalent to either $exp_a$ or $exp_s$ form continuous sets of edges along the path such that $exp_a$ and $exp_s$ can be translated along those sets into a common scope.  

Note that variables declared by a program as accessible across different functions such as global variables in C/C++ do not require the translation, although the use of such kind of variables is not very common. We refer to both function parameters and such kind of variables collectively as nonlocal variables. And we refer to an expression consisting of only nonlocal variables as a nonlocal expression.

The low-level implementation of expression translation in Senx consists of two functions.  Function \texttt{translate\_se\_to\_scopes}, listed in Algorithm~\ref{alg:translate_se_to_scopes}, is the core of expression translation. It translates a particular expression $expr$ to the scope of all candidate functions along the call stack $stack$.  

\begin{algorithm}
        \caption{Translating an expression to the scope of each function on the call stack.}
        \label{alg:translate_se_to_scopes}
        \begin{algorithmic}
		\Require $stack$: a call stack consists of an ordered list of call instruction \\ \hspace*{\algorithmicindent} $expr$: the expression to be translated \\ \hspace*{\algorithmicindent} $inst$: the instruction to which $expr$ is associated
		\Ensure $translated\_exprs$: the translated $expr$ in the scope of each caller function on the call stack
        \begin{algorithmic}[1]
                \Procedure{translate\_se\_to\_scopes}{}
			\LineComment{Translate $expr$ to an expression in which all the variables are the parameters of $func$}
			\State $func\gets$get\_func($inst$)
      			\State $expr\gets$make\_nonlocal\_expr($func,inst,expr$)
			\If{$expr \ne \varnothing$}
			\For{$call \in stack$}
				\LineComment{Substitute each parameter variable in $expr$ with its correspondent argument used in $call$}
				\State $expr\gets$substitute\_parms\_with\_args($call,expr$)
				\State $func\gets$get\_func($call$)
				\State $translated\_exprs[func]\gets expr$
               			\State $expr\gets$make\_nonlocal\_expr($func,call,expr$)
				\If{$expr=\varnothing$}
				\Break
				\EndIf
			\EndFor
			\EndIf
                	\State \textbf{return} ${translated\_exprs}$
                \EndProcedure
        \end{algorithmic}
	\end{algorithmic}
\end{algorithm}

\begin{algorithm}
        \caption{Making a nonlocal expression.}
        \label{alg:make_nonlocal_expr}
        \begin{algorithmic}
		\Require $f$: a function \\ \hspace*{\algorithmicindent}	$inst$: an instruction in $f$ \\ \hspace*{\algorithmicindent}	$expr$: the RHS expression associated with $inst$
		\Ensure $nonlocal\_expr$: the nonlocalized $expr$
        \begin{algorithmic}[1]
                \Procedure{make\_nonlocal\_expr}{}
					 \LineComment{$mapping$ stores the nonlocal definition for each variable within $expr$}
					 \State $mapping \gets \emptyset$
					 \For{$var \in expr$}
					 \If{$\neg$ is\_var\_nonlocal($f, var$)}
					 \State $def \gets$find\_nonlocal\_def\_for\_var($f, inst, var$)
					 \If{$def=\varnothing$}
					 \LineComment{We cannot find a nonlocal definition for $var$}
					 \State \textbf{return} $\varnothing$
					 \Else
					 \State $mapping[var] \gets def$
				    \EndIf
					 \EndIf
					 \EndFor
					 \LineComment{Substitute the occurrence of each variable with its nonlocal definition}
					 \State $nonlocal\_expr \gets$substitute\_vars($expr, mapping$)
                \State \textbf{return} ${nonlocal\_expr}$
                \EndProcedure
        \end{algorithmic}
	\end{algorithmic}
\end{algorithm}

We illustrate how it works with the code in Listing~\ref{lst:example_vulnerability}. For simplicity, we use source code line numbers to represent the corresponding instructions. To translate the buffer size involved in the buffer overflow, Senx finds that the buffer is allocated from a call to \texttt{malloc} at line~\ref{lst:malloc_call} from the call stack that it associates with each memory allocation, and invokes \texttt{translate\_se\_to\_scopes} with $stack=$[line~\ref{lst:foo_malloc_call} ], $expr=$``\texttt{x*y+1}'', $inst=$line~\ref{lst:malloc_call} , $func=\texttt{foo\_malloc}$. The function first converts ``\texttt{x*y+1}'' into a definition in which variables are all parameters of \texttt{foo\_malloc}, which we call a nonlocal definition, if such conversion is possible. This conversion is done by function \texttt{make\_nonlocal\_expr} listed in Algorithm~\ref{alg:make_nonlocal_expr}, which tries to find a nonlocal definition for each variable in $expr$ and then substitutes each variable with its matching nonlocal definition. \texttt{make\_nonlocal\_expr} relies on \texttt{find\_nonlocal\_def\_for\_var}, which recursively finds reaching definitions for local variables in a function, eventually building a definition for them in terms of the function parameters, global variables or the return values from function calls.  Note that a nonlocal definition can only be in the form of an arithmetic expression without involving any functions. In this case, the resulting $expr$ is also ``\texttt{x*y+1}'' because both \texttt{x} and \texttt{y} are parameters of \texttt{foo\_malloc}. 

It then iterates each call instruction in $stack$, starting from line~\ref{lst:foo_malloc_call}. For each call instruction, it substitutes the parameters in $expr$ with the arguments used in the call instruction. For line~\ref{lst:foo_malloc_call}, it substitutes \texttt{x} with \texttt{rows} and \texttt{y} with \texttt{cols+1}, respectively, by calling helper function \texttt{substitute\_parms\_with\_args}. As a consequence, ``\texttt{x*y+1}'' becomes ``\texttt{rows*(cols+1)+1}''. Hence it associates ``\texttt{rows*(cols+1)+1}'' with function \texttt{foo} and stores the association in $expr\_translated$, because line~\ref{lst:foo_malloc_call} exists in function \texttt{foo}. After that, it tries to convert ``\texttt{rows*(cols+1)+1}'' into a nonlocal definition with respect to \texttt{foo}. At this point, it halts because both \texttt{rows} and \texttt{cols} are assigned with return values of calls to function \texttt{extract\_int}. Otherwise, it will move on to the next function on the call stack and continue the translation upwards the call stack.  However, in this case, expression translation is also able to translate the memory access range expression from the scope of \texttt{bar} into the scope of \texttt{foo}.  Thus,  Senx places the patch predicate in \texttt{foo}.  If expression translation fails to converge the expression, Senx will abort patch generation.

\section{Implementation}\label{sec:implementation}

We have implemented Senx as an extension of the KLEE LLVM execution engine~\cite{KLEE}. Like KLEE, Senx works on C/C++ programs that are compiled into LLVM bitcode \cite{LLVM}. 

We re-use the LLVM bitcode execution portion of KLEE, and as described in Section~\ref{sec:symbolic_expression_builder}, to implement our expression builder, but do not use any of the constraint collection or solving parts of KLEE.  For simplicity and ease of debugging, we represent our expressions as text strings. To support arithmetic operations and simple math functions on expressions, we leverage {GiNaC}, a C++ library designed to provide support for symbolic manipulations of algebra expressions \cite{GiNac}.

We implement a separate LLVM transformation pass to annotate LLVM bitcode with information on loops such as the label for loop pre-header and header, which is subsequently used by access range analysis. This pass relies on LLVM's canonicalization of natural loops to normalize loops \cite{LLVM_LOOP_CANONICALIZATION}. We extend LLVMSlicer~\cite{LLVMSlicer} for loop cloning. To locate error handling code, we use Talos \cite{Talos}. 

Our memory allocation logger uses KLEE to interpose on memory allocations and stores the call stack for each memory allocation. Senx extends KLEE to detect integer overflows and incorporates the existing memory fault detection in KLEE to trigger our patch generation.  For alias analysis, Senx leverages DSA pointer analysis~\cite{DSA}.


Senx is implemented with 2,543 lines of C/C++ source code, not including the Talos component used to identify error handling code. Half of the source code is used to implement expression builder, which forms the foundation of other components of Senx. 

\section{Evaluation}\label{sec:senx_evaluation}
First, we evaluate the effectiveness of Senx in fixing real-world vulnerabilities. Second, we manually examine the produced patches for correctness and compare them to the developer created patch.  For the sake of space, we only describe two of the patches in detail. Last, we measure the applicability of loop cloning, access range analysis, and expression translation using a larger dataset.


\begin{table} [tb]
	\caption{Applications for testing real-world vulnerabilities.} \label{tbl:programs_with_vulnerabilities}
	\begin{center}
		\begin{tabular}{|l|p{1.9in}|r|}
			\hline
			\textbf{App.} & \textbf{Description} & \textbf{SLOC} \\
			\hline
			autotrace & a tool to convert bitmap to vector graphics & 19,383 \\
			binutils & a collection of programming tools for managing and creating binary programs & 2,394,750 \\
			libming & a library for creating Adobe Flash files & 88,279 \\
			libtiff & a library for manipulating TIFF graphic files & 71,434 \\
			PHP & the official interpretor for PHP programming language & 746,390 \\
			sqlite & a relational database engine & 189,747 \\
			ytnef & TNEF stream reader & 15,512 \\
			zziplib & a library for reading ZIP archives & 24,886 \\
			jasper & a codec for JPEG standards & 30,915 \\
			libarchive & a multi-format archive and compression library & 158,017 \\
			potrace & a tool for tracing bitmap graphics & 20,512 \\
			\hline
			\textbf{Total} & N/A & 3,817,268 \\
			\hline 
		\end{tabular}
		\vspace{-5pt}
	\end{center}
\end{table}

\subsection{Experiment Setup}\label{sec:eval_setup}

We build a corpus of vulnerabilities for Senx to attempt to patch by searching online vulnerability databases ~\cite{NVD, MITRE, CVEdetails}, software bug report databases, developers' mailing groups~\cite{PHPBugs, RedHatBugzilla, coreutilsBugs}, and exploit databases~\cite{exploitdb}.  We focus on vulnerabilities that fall into one of the three types of vulnerabilities Senx can currently handle.  We then select vulnerabilities that meet the following three criteria: 1) an input to trigger the vulnerability is either available or can be created from the information available, 2) the vulnerable application can be compiled into LLVM bitcode and executed correctly by KLEE, and 3) the vulnerable application uses \texttt{malloc} to allocate memory as Senx currently relies on this to infer the allocation size of objects.  Applications that use custom memory allocation routines are currently not supported by Senx.  We obtain the vulnerability-triggering inputs or information about such inputs from the blogs of security researchers, bug reports, exploit databases, mailing groups for software users, or test cases attached to patch commits~\cite{agostinoblog, agostinopoc, exploitdb, sourceware, maptools, sqlite-users}.

From this, we construct a corpus of 42 real-world buffer overflow, integer overflow and bad cast vulnerabilities along with proof of concept exploits to evaluate the effectiveness of Senx in patching vulnerabilities. The vulnerabilities are drawn from 11 applications show in Table~\ref{tbl:programs_with_vulnerabilities}, which include 8 media and archive tools and libraries, PHP, sqlite, and a collection of programming tools for managing and creating binary programs.  The associated vulnerabilities consist of 19 buffer overflows, 13 integer overflows, and 10 bad cast vulnerabilities.

%

%


All our experiments were conducted on a desktop with quad-core 3.40GHz Intel i7-3770 CPU, 16GB RAM, 3TB SATA hard drive and 64-bit Ubuntu 14.04.

\newcommand{\NA}{---}
\begin{table*}[t]
	\footnotesize
	\caption{Patch generation by Senx}\vspace{-10pt}
	  \label{tbl:senx-vulnerabilities}
  \centering
  \begin{tabular}{|l|l|l|c|c|c|c|c|}
	\hline
    \textbf{App.} & \textbf{CVE\#} & \textbf{Type} & \textbf{Expressions} & \textbf{Loop Analysis} & \textbf{Patch Placement} & \textbf{Data Access} &  \textbf{Patched?} \\
	 \hline
	sqlite & \textcolor{red}{CVE-2013-7443} & \textcircled{3} & Determinate & \NA & \textcolor{red}{Failed} & \NA & \ding{55} \\
& CVE-2017-13685 & \textcircled{3} & Determinate & \NA & Trivial & Simple & \ding{51} \\
\hline
zziplib & CVE-2017-5976 & \textcircled{1} & Determinate & Cloned & Translated & Complex & \ding{51} \\
& CVE-2017-5974 & \textcircled{3} & Determinate & \NA & Translated & Complex & \ding{51} \\
& CVE-2017-5975 & \textcircled{3} & Determinate & \NA & Translated & Complex & \ding{51} \\
\hline
Potrace & CVE-2013-7437 & \textcircled{2} & Determinate & \NA & Trivial & Complex & \ding{51} \\
\hline
libming & CVE-2016-9264 & \textcircled{3} & Determinate & \NA & Trivial & Simple & \ding{51} \\
\hline
libtiff & \textcolor{red}{CVE-2016-9273} & \textcircled{1} & \textcolor{red}{Indeterminate} & \NA & \NA & \NA & \ding{55} \\
& CVE-2016-9532 & \textcircled{1} & Determinate & Cloned & Trivial & Complex & \ding{51} \\
& CVE-2017-5225 & \textcircled{1} & Determinate & ARA & Trivial & Simple & \ding{51} \\
& CVE-2016-10272 & \textcircled{1} & Determinate & ARA & Translated & Simple & \ding{51} \\
& CVE-2016-10092 & \textcircled{3} & Determinate & \NA & Translated & Simple & \ding{51} \\
& CVE-2016-5102 & \textcircled{3} & Determinate & \NA & Trivial & Simple & \ding{51} \\
& CVE-2006-2025 & \textcircled{2} & Determinate & \NA & Trivial & Complex & \ding{51} \\
\hline
libarchive & CVE-2016-5844 & \textcircled{2} & Determinate & \NA & Trivial & Complex & \ding{51} \\
\hline
jasper & CVE-2016-9387 & \textcircled{2} & Determinate & \NA & Trivial & Complex & \ding{51} \\
& CVE-2016-9557 & \textcircled{2} & Determinate & \NA & Trivial & Complex & \ding{51} \\
& \textcolor{red}{CVE-2017-5501} & \textcircled{2} & Determinate & \NA & \textcolor{red}{Failed/Error handling} & \NA & \ding{55} \\
\hline
ytnef & CVE-2017-9471 & \textcircled{1} & Determinate & Cloned & Trivial & Simple & \ding{51} \\
& CVE-2017-9472 & \textcircled{1} & Determinate & Cloned & Trivial & Simple & \ding{51} \\
& \textcolor{red}{CVE-2017-9474} & \textcircled{1} & Determinate & \textcolor{red}{Failed} & \NA & \NA & \ding{55} \\
\hline
PHP & CVE-2011-1938 & \textcircled{1} & Determinate & ARA & Translated & Simple & \ding{51} \\
& CVE-2014-3670 & \textcircled{1} & Determinate & ARA & Translated & Complex & \ding{51} \\
& CVE-2014-8626 & \textcircled{1} & Determinate & Cloned & Trivial & Simple & \ding{51} \\
\hline
binutils & CVE-2017-15020 & \textcircled{1} & Determinate & ARA & Translated & Simple & \ding{51} \\
& CVE-2017-9747 & \textcircled{1} & Determinate & Cloned & Translated & Simple & \ding{51} \\
& CVE-2017-12799 & \textcircled{3} & Determinate & \NA & Trivial & Simple & \ding{51} \\
& \textcolor{red}{CVE-2017-6965} & \textcircled{3} & Determinate & \NA & \textcolor{red}{Failed} & \NA & \ding{55} \\
& CVE-2017-9752 & \textcircled{3} & Determinate & \NA & Translated & Simple & \ding{51} \\
& \textcolor{red}{CVE-2017-14745} & \textcircled{2} & Determinate & \NA & \textcolor{red}{Failed} & \NA & \ding{55} \\
\hline
autotrace & \textcolor{red}{CVE-2017-9151} & \textcircled{1} & \textcolor{red}{Indeterminate} & \NA & \NA & \NA & \ding{55} \\
& \textcolor{red}{CVE-2017-9153} & \textcircled{1} & \textcolor{red}{Indeterminate} & \NA & \NA & \NA & \ding{55} \\
& CVE-2017-9156 & \textcircled{1} & Determinate & ARA & Trivial & Simple & \ding{51} \\
& CVE-2017-9157 & \textcircled{1} & Determinate & ARA & Trivial & Simple & \ding{51} \\
& \textcolor{red}{CVE-2017-9168} & \textcircled{1} & Determinate & \textcolor{red}{Failed} & \NA & \NA & \ding{55} \\
& \textcolor{red}{CVE-2017-9191} & \textcircled{1} & Determinate & ARA & \textcolor{red}{Failed} & \NA & \ding{55} \\
& CVE-2017-9161 & \textcircled{2} & Determinate & \NA & Trivial & Simple & \ding{51} \\
& CVE-2017-9183 & \textcircled{2} & Determinate & \NA & Trivial & Complex & \ding{51} \\
& CVE-2017-9197 & \textcircled{2} & Determinate & \NA & Trivial & Complex & \ding{51} \\
& CVE-2017-9198 & \textcircled{2} & Determinate & \NA & Trivial & Complex & \ding{51} \\
& CVE-2017-9199 & \textcircled{2} & Determinate & \NA & Trivial & Complex & \ding{51} \\
& CVE-2017-9200 & \textcircled{2} & Determinate & \NA & Trivial & Complex & \ding{51} \\
\hline

	\hline
  \end{tabular}

\end{table*}

\subsection{How Effective is Senx in Patching Vulnerabilities?}

For each vulnerability of an application, we run the corresponding program under Senx with a vulnerability-triggering input. If Senx generates a patch, we examine the patch for correctness.  To determine if a patch is correct, we apply the three following tests a) we check for semantic equivalence with the official patch released by the vendor if available and semantic correctness by analyzing the code, b) we apply the patch and verify that the vulnerability is no longer triggered by the input and c) we check as best we can that the patch does not interfere with regular operation of the application by using the application to process benign inputs. If Senx aborts patch generation, we examine what caused Senx to abort.



Our results are summarized in Table~\ref{tbl:senx-vulnerabilities}. Column ``Type'' indicates whether the vulnerability is a \textcircled{1} buffer overflow, \textcircled{2} bad cast, or \textcircled{3} integer overflow. Column ``Expressions'' shows whether Senx can successfully construct all expressions that are required to synthesize a patch, as some code constructs may contain expressions outside of the theories Senx supports in its symbolic ISA.  ``Loop Analysis'' describes whether 
loop cloning or access range analysis (ARA) is used if the vulnerability contained a loop. ``Patch Placement'' lists the type of patch placement: ``Trivial'' means that the patch is placed in the same function as the vulnerability and ``Translated'' means that the patch must be translated to a different function.   ``Data Access'' describes whether or not the patch predicate involves complex data access such as fields in a struct or array indices.  Finally, ``Patched?'' summarizes whether the patch generated by Senx fixes a vulnerability. The 10 vulnerabilities where Senx aborts generating a patch are highlighted in red.

Over the 42 vulnerabilities, Senx generates 32 (76.2\%) patches, all of which are correct according to our three criteria.  Of the 14 patched buffer overflows, loop analysis is roughly split between loop cloning and access range analysis (6 and 8 respectively).  Senx elects not to use loop cloning mainly due to two causes.  First, due to an imprecise alias analysis that does not distinguish different fields of structs correctly, the program slicing tool utilized by Senx may include instructions that are irrelevant to computing loop iterations into slices.  Unfortunately these instructions call functions that can have side-effects so the slices cannot be used by Senx. Second, for a few cases the entire body of the loops is control dependent on the result of a call to a function that has side-effects. For example, the loops involved in CVE-2017-5225 are only executed when a call to \texttt{malloc} succeeds. Because \texttt{malloc} can make system calls, Senx also cannot clone the loops.
 
Senx must place 23.8\% of the patches in a function different from where the vulnerability exists.  This is particularly acute for buffer overflows (46.2\% of cases), which have to compare a buffer allocation with a memory access range.  This illustrates that expression translation contributes significantly to the patch generation ability of Senx, particularly for buffer overflows, which make up the majority of memory corruption vulnerabilities.  Senx's handling of complex data accesses is also used in 48.5\% of the patches, indicating this capability is required to handle a good number of vulnerabilities

Senx aborts patch generation for 10 vulnerabilities. The dominant cause for these aborts is that Senx is not able to converge to a function scope where all symbolic variables in the patch predicate are available.  There is also one case (jasper-CVE-2017-5501) where Senx cannot find appropriate error-handling code to synthesize the patch. In these cases, the patch requires more significant changes to the application code that are beyond the capabilities of Senx. In other cases, Senx detects that there are multiple reaching definitions for patch predicates that it does not have an execution input for.  Currently, Senx only accepts one execution path executed by the single vulnerability-triggering input.  In the future we plan to handle these cases by allowing Senx to accept multiple inputs to cover the paths along which the other reaching definitions exist.  Finally, Senx aborts for a couple of vulnerabilities because both loop cloning and access range analysis fail.



\subsection{Patch Case Study}
Out of the 32 generated patches, we select 2 patches to describe in detail.

\scbf{libtiff-CVE-2017-5225} This is a heap buffer overflow in libtiff, which can be exploited via a specially crafted TIFF image file. The overflow occurs in a function \texttt{cpContig2SeparateByRow} that parses a TIFF image into rows and dynamically allocates a buffer to hold the parsed image based on the number of pixels per row and bits per pixel.  By using an inconsistent bits per pixel parameter, the attacker can cause libtiff to allocate a buffer smaller than the size of the pixel data and cause a buffer overflow.

When Senx captures the buffer overflow via running libtiff with a crafted TIFF image file, it first identifies that the buffer is allocated using the value of variable \texttt{scanlinesizein} and the starting address of the buffer is stored in variable \texttt{inbuf}. Hence it uses [\texttt{inbuf}, \texttt{inbuf + scanlinesizein}] to denote the buffer range. Senx then finds that the buffer overflow occurs in a 3-level nested loop and that the pointer used to access the buffer is dependent on the loop induction variable.  Senx classifies the vulnerability as a buffer overflow.

Loop cloning fails because the loop slice is dependent on a call to \texttt{\_TIFFmalloc}, which subsequently calls \texttt{malloc}.  Thus, Senx applies access range analysis.  Access range analysis detects that only the outer and inner-most loops affect the memory access pointer and from the extracted induction variables, computes the expression [\texttt{inbuf}, \texttt{inbuf+spp*imagewidth}] to represent the access range.  

Because both the buffer range and the access range start at \texttt{inbuf}, Senx synthesizes the patch predicate as \texttt{spp*imagewidth > scanlinesizein}. Senx then finds that \texttt{cpContig2SeparateByRow} contains error handling code, which has a label \texttt{bad}, and generates the patch as below. As the buffer allocation and overflow occur in the same function, Senx puts the patch immediately before the buffer allocation.

\begin{lstlisting}[floatplacement=F,language=C,numbers=none]
	if (spp*imagewidth > scanlinesizein) 
		goto bad;
\end{lstlisting}

The official patch invokes the same error handling and is placed at the same location as Senx's patch.  However, the official patch checks that ``\texttt{(bps != 8)}''.  From further analysis, we find that both patches are equivalent, though the human-generated patch relies on the semantics of the libtiff format, while Senx's patch directly checks that the loop cannot exceed the size of the allocated buffer.


\scbf{libarchive-CVE-2016-5844} This integer overflow in the ISO parser in libarchive can result in a denial of service via a specially crafted ISO file. The overflow happens in function \texttt{choose\_volume} when it multiplies a block index, which is a 32-bit integer, with a constant number. This can exceed the maximum value that can be represented by a 32-bit integer and overflow into a negative number, which is then used as a file offset.

Senx detects the integer overflow when it runs libarchive's ISO parser with a crafted ISO file. It generates an expression of the overflown value as the product of 2048 and  \texttt{vd$\rightarrow$location}.  Further Senx detects that the overflown value is assigned to a 64-bit variable \texttt{skipsize}, thus classifying this as a repairable integer overflow.  Senx patches the vulnerability by casting the 32-bit value to a 64-bit value before multiplying:


\begin{lstlisting}[floatplacement=F,language=C,numbers=none]
	- skipsize = LOGICAL_BLOCK_SIZE * vd->location;
	+ skipsize = 2048 * (int64_t)vd->location;
\end{lstlisting}

The official patch is essentially identical to the patch generated by Senx. The only difference is that the official patch uses the constant \texttt{LOGICAL\_BLOCK\_SIZE} rather than its equivalent value 2048 in the multiplication.

\subsection{Applicability}

We evaluate how applicable loop cloning, access range analysis and expression translation are across a larger dataset.  To generate such a dataset, we extract all loops that access memory buffers and the allocations of these buffers from the 11 programs in GNU Coreutils, regardless of whether they contain vulnerabilities or not.  We then apply Senx's loop analysis to all loops and find that loop cloning can be applied to 88\% of the loops and access range analysis can be applied to 46\% of the loops.  This is in line with our results from the vulnerabilities. For the sake of space, we describe the details of these experiments in the Appendix.

\section{Related Work}\label{sec:related_work}
\subsection{Automatic Patch Generation}
\scbf{Leveraging Fix Patterns} By observing common human-developer generated patches, PAR generates patches using fix patterns such as altering method parameters, adding a null checker, calling another method with the same arguments, and adding an array bound checker~\cite{PAR}. Senx differs from PAR in two aspects. First, PAR is unable to generate a patch when the correct variables or methods needed to synthesize a patch are not accessible at the faulty function or method. Second, PAR uses a trial-and-error approach that tries out not only each fixing pattern upon a given bug, but also variables or methods that are accessible at the faulty function or method to synthesize a patch. On the contrary, Senx employs a guided approach that identifies the type of the given bug and chooses a corresponding patch model to generate the patch for the bug and systematically finds the correct variables to synthesize the patch based on semantic information provided by a patch model. 


LeakFix fixes memory leak bugs by adding a memory deallocation statement for a leaked memory allocation at the correct program location~\cite{LeakFix}. By abstracting a program into a CFG containing only program statements related to memory allocation, deallocation, and usage, LeakFix transforms the problem of finding a fix for a memory leak into searching for a CFG edge where a memory deallocation statement can be added to fix the memory leak without introducing new bugs.


\scbf{Using Program Mutations}
GenProg is a pioneering work that induces program mutations, i.e. genetic programming, to generate patches~\cite{GenProg}. Leveraging test suites, it focuses on program code that is executed for negative test cases but not for positive test cases and utilizes program mutations to produce modifications to a program. As a feedback to its program mutation algorithm, it considers the weighted sum of the positive test cases and negative test cases that the modified program passes. Treating all the results of program mutations as a search space, its successor improves the scalability by changing to use patches instead of abstract syntax trees to represent modifications and exploiting search space parallelism~\cite{Goues2012-study}.


\scbf{Using SMT Solvers}
SemFix~\cite{SemFix} and Angelix~\cite{Angelix} use constraint solving to find the needed expression to replace incorrect expressions used in a program. By executing a target program symbolically with both inputs triggering a defect and inputs not triggering the defect, they identify the constraints that the target program must satisfy to process both kinds of inputs correctly. They then synthesize a patch using component-based program synthesis, which combines components such as variables, constants, and arithmetic operations to synthesize a symbolic expression that can make the target program satisfy the identified constraints. 


\scbf{Learning from Correct Code}
Prophet learns from existing correct patches~\cite{Prophet}. It uses a parameterized log-linear probabilistic model on two features extracted from the abstract trees of each patch: 1) the way the patch modifies the original program and 2) the relationships between how the values accessed by the patch are used by the original program and by the patched program. With the probabilistic model, it ranks candidate patches that it generated for a defect by the probabilities of their correctness. Finally it uses test suites to test correctness of the candidate patches. Like other generate-and-validate automatic patch generation techniques, its effectiveness depends on the quality of the test suites.



\subsection{Prevention of Exploiting Security Vulnerabilities}
\scbf{Fortifying Programs} One way to prevent vulnerabilities from being exploited is by fortifying programs to make them more robust to malicious inputs. Software Fault Isolation (SFI) instruments checks before memory operations to ensure that they cannot corrupt memory~\cite{sfi, native, rocksalt, LMP}. Control Flow Integrity (CFI) learns valid control flow transfers of a program and validates control flow transfers to prevent execution of exploit code~\cite{Zhang2013CFICOTS, Criswell2014, Tice2014}. Alternatively, some techniques modify the layout or permissions of critical memory regions to detect or prevent exploits~\cite{stackguard, stackshield, Kruiser, shacham04, pax, Chameleon}.

By contrast, Talos introduces the notion of Security Workarounds for Rapid Response (SWRR), which steers program execution away from a vulnerability to error handling code, and instruments SWRRs to programs to prevent malicious inputs from triggering vulnerabilities~\cite{Talos}. 

While these techniques prevent exploit of vulnerabilities at the cost of extra runtime overhead or loss of program functionality, Senx generates patches to fix vulnerabilities without imposing such a cost.

\scbf{Filtering Inputs} Some techniques detect and simply filter out malicious inputs~\cite{Costa2007, Long2014-InputFilter, Susskraut2007, Wang2004, VulnerabilitySignature}. Among them, Bouncer combines static analysis and symbolic analysis to infer the constraints to exploit a vulnerability and generates an input filter to drop such malicious inputs~\cite{Costa2007}. Shields models a vulnerability as a protocol state machine and constructs network filters based on it~\cite{Wang2004}. 

While most of these techniques identify malicious inputs by the semantics or syntax of the malicious inputs, some of them focus on the semantics of vulnerabilities~\cite{Long2014-InputFilter, VulnerabilitySignature}. Similar to these techniques, Senx also uses the semantics of vulnerabilities to synthesize patches. However, Senx has a different goal of generating patches to fix vulnerabilities.

\scbf{Rectifying Inputs} Alternative to filtering inputs, some techniques rectify malicious inputs to prevent them from triggering vulnerabilities. With taint analysis, SOAP learns constraints on input by observing program executions with benign inputs. From the constraints that it has learned, it identifies input that violates the constraints and tries to change the input to make it satisfy the constraints. By doing so, it not only renders the input harmless but also allows the desired data in the rectified input to be correctly processed~\cite{Long2012}.

Based on the observation that exploit code embedded in inputs is often fragile to any slight changes, A2C encodes inputs with a one-time dictionary and decodes them only when the program execution goes beyond the paths likely to have vulnerabilities in order to disable the embedded exploit code~\cite{A2C}.

\section{Conclusion}\label{sec:senx_conclusion}
This paper presents the design and implementation of Senx, a system that automatically generates patches for buffer overflow, bad cast, and integer overflow vulnerabilities. Senx uses three novel program analysis techniques introduced in this paper: loop cloning, access range analysis and expression translation.  In addition, Senx utilizes an expression representation that facilitates the translation of expressions extracted from symbolic execution back into C/C++ source code.  Enabled by these techniques, Senx generates patches correctly for 76\% of the 42 real-world vulnerabilities.  Senx's main limitations are limited precision in alias analysis for loop cloning, and the inability to converge expressions to find a location in the program where all necessary variables are in scope.  


\bibliographystyle{ACM-Reference-Format}
\bibliography{bibfile}

\section*{Appendix}\label{sec:appendix}

We measure how often expression translation is able to converge the memory access range and buffer allocation size into a single function scope, and find that it is able to do so in 85\% of the cases. 

We use 11 programs from the GNU Coreutils as listed in Table~\ref{tbl:programs} to evaluate the applicability of our analysis techniques.  The most common reasons for Senx's access range analysis to be aborted is that loops cannot be normalized by LLVM.  For example, the number of times a loop that parses string input iterates depends on the content of the string.  Such a string cannot be symbolically analyzed by access range analysis.

To understand the reasons that can cause expression translation to abort, we try to converge the buffer size and memory access range for the loops that we could successfully analyze and tabulate the results in Table~\ref{tbl:eval_symbolic_expression_translation}.  The ``Access Range'' column tabulates the average percentage of functions in the loop's call stack that expression translation could translate the memory access range into and ``Buffer Range'' tabulates the average percentage of functions in the buffer allocation's call stack that expression translation could translate the buffer allocation size into.  Finally ``Converged'' indicates out of all loops, what percentage could expression translation find a  common function scope in which to place the patch.  As we can see, it seems that the buffer allocation size frequently takes parameters that are calculated fairly close in the call stack to the allocation point, and those values are not available higher up in the call chain, thus limiting the functions scopes many of these cases could be converged to.

\begin{table}
		\caption{Programs for evaluating applicability.}
	\label{tbl:programs}
	\begin{center}
		\begin{tabular}{|l|l|r|r|}
			\hline
			\textbf{Program} & \textbf{Type} & \textbf{SLOC} & \textbf{LLVM bitcode} \\
			\hline
			sha512sum & data checksum & 581 & 135KB \\
			pr & text formatting & 1,723 & 194KB \\
			head & text manipulation & 761 & 109KB \\
			dir & directory listing & 3,388 & 418KB \\
			od & file dumping & 1,368 & 237KB \\
			ls & directory listing & 3,388 & 418KB \\
			base64 & data encoding & 238 & 91KB \\
			wc & text processing & 784 & 120KB \\
			cat & file concatenating & 495 & 182KB \\
			sort & data sorting & 3,251 & 433KB \\
			printf & format and print data & 694 & 198KB \\
			\hline
			\textbf{AVG} & N/A & 1,516 & 230KB \\
			\hline 
		\end{tabular}
	\end{center}
\end{table}

\begin{table}
			\caption{Convergence of expression translation.}
	\label{tbl:eval_symbolic_expression_translation}	
	\begin{center}
		\begin{tabular}{|l|r|r|r|}
			\hline
			\textbf{Program} & \textbf{Access Range} & \textbf{Buffer Range} & \textbf{Converged} \\
			\hline
			pr & 100\% & 10\% & 100\%  \\
			head & 100\% & 25\% & 100\%   \\
			tr & 86\% & 36\% & 100\%  \\
			od & 54\% & 16\% & 58\%  \\
			cat & 100\% & 33\% & 100\%  \\
			dir & 71\% & 14\% & 57\%  \\
			ls & 42\% & 33\% & 34\%  \\
			base64 & 100\% & 33\% & 100\%  \\
			md5sum & 100\% & 33\% & 100\%  \\
			sha512sum & 97\% & 80\% & 97\%  \\
			sort & 91\% & 10\% & 90\%  \\
			\hline
			\textbf{AVG.} & 85\% & 29\% & 85\%  \\
			\hline 
		\end{tabular}
	\end{center}
\end{table}

\end{document}